\documentclass[11pt,epsf,epsfig,psfrag]{article}
\usepackage{hyperref}
\usepackage[bf]{caption}
\usepackage{epsfig}
\usepackage{psfrag}

\usepackage{float}
\usepackage{graphicx}
\usepackage{epstopdf}
\usepackage{subfig}
\def\ga{\mathrel{\raise.3ex\hbox{$>$\kern-.75em\lower1ex\hbox{$\sim$}}}}
\def\la{\mathrel{\raise.3ex\hbox{$<$\kern-.75em\lower1ex\hbox{$\sim$}}}}

\def\I_M{{I_{\scriptscriptstyle M\times M}}}

\begin{document}

\thispagestyle{empty}

\vskip 2cm 


\begin{center}
{\Large \bf A Comparative Note on Tunneling in AdS and \\
\vskip 0.2cm
in its Boundary Matrix Dual}
\end{center}

\vskip .2cm

\vskip 1.2cm

\centerline{ \bf B. Chandrasekhar$^1$\footnote{bcsekhar@fma.if.usp.br}
, Sudipta Mukherji$^2$\footnote{mukherji@iopb.res.in}, Anurag Sahay$^2$\footnote{anurag@iopb.res.in} 
and Swarnendu 
Sarkar$^3$\footnote{ssarkar@physics.du.ac.in}}

\vskip 7mm \centerline{ \it $^1$ Departamento de Fisica Mathematica,
Universidad de Sao Paulo, Brazil 05315970}

\vskip 4mm\centerline{ \it $^2$ Institute of Physics, 
Bhubaneswar-751 005, India}

\vskip 4mm\centerline{ \it $^3$ Department of Physics and Astrophysics, University of Delhi, Delhi 110007, 
India}

\vskip 1.2cm
\vskip 1.2cm
\centerline{\bf Abstract}
\noindent

For charged black hole, within the grand canonical ensemble, the decay rate from thermal AdS 
to the black hole at a fixed high temperature increases with the  chemical potential.
We check that this feature is well captured by a phenomenological matrix model 
expected to describe its strongly coupled dual. This comparison is made by explicitly 
constructing the kink and bounce solutions around the de-confinement transition 
and evaluating the matrix model effective potential on the solutions.
\newpage
\setcounter{footnote}{0}
\noindent

\section{Introduction and Summary}

Quite a while back, Hawking and Page (HP) argued that at temperature below
 $T_0  = {\sqrt 2}/(\pi l)$, thermal AdS background\footnote{In our
notation $l$ is related to
the AdS scale. Later, we work with units such that $l = 1$.}
with a non-contractible Euclidean time circle $\beta = 
1/T$, is
stable against collapse to form a black hole\cite{Hawking:1982dh}. However, at a temperature higher than $T_0$, two 
Schwarzschild black 
holes nucleate, which  following the literature, we call  small and big black holes. While 
the small black hole is locally unstable, the other one has a positive specific heat. In fact, at a temperature $T$ 
greater
than $T_1 = 3/(\pi l) > T_0$, the large black hole dominates the canonical ensemble for quantum gravity defined as a path 
integral over the metric and all other fields asymptotic to AdS with the periodicity of the time direction being 
$\beta$.
It is believed that the small unstable black hole facilitates this transition. 
In the dual gauge theory, this represents a de-confining transition
for ${\cal N} =4$, large $N$, ~$SU(N)$ strongly coupled Yang-Mills on $S^3$. Though for gauge theory at
finite coupling, there is no systematic way to compute an effective action, one 
may hope to construct a phenomenological model which captures this transition. Indeed in \cite{AlvarezGaume:2005fv},
partition function of YM theory was expressed as  a matrix integral over the effective action 
involving Wilson-Polyakov loop. This model, which we review shortly, also predicts 
certain qualitative features that are not visible in the bulk supergravity approximation.
Our aim in this paper is to put this matrix model to further qualitative checks. This is described in the
following paragraph.

As we discussed in the beginning, HP transition is expected to be facilitated by the small unstable black hole. 
A particularly simple way to qualitatively analyze this transition would be to construct a {\it bulk effective}
potential and to find instanton and bounce solutions at and above the HP temperature. 
In this regard, it turns out that the Bragg-Williams(BW) prescription 
is a particularly suitable framework (see 
\cite{Banerjee:2010ve} for details). In this approach, one constructs
an approximate expression for the off-shell free energy function in terms of the order parameter and
uses the condition that its equilibrium value minimizes the free energy. For  describing
HP transition from thermal AdS to the big black hole, horizon radius is an order parameter
of the effective potential and the temperature then acts as an external tunable parameter.
At the saddle point of this potential, however, the temperature gets related with the horizon radius in a way that is 
consistent with black hole thermodynamics. As we will describe shortly, 
following Coleman's prescription \cite{Coleman:1977py}, it is particularly simple 
to construct an exact instanton  at HP temperature $T = T_1$ and the bounce 
above $T_1$ -- a region where the big black hole becomes globally stable. 
Subsequently, one can compute the semi-classical decay rate from one phase to the other.
Introduction of electrical charge into the system makes the 
chemical potential non-zero. The black hole in question is now the Reissner-Nordstr\"om
black hole. In the grand canonical ensemble, as it was pointed out
in \cite{Chamblin:1999tk}, provided the chemical potential is less than a certain critical
value, HP transition still persists. However, the critical temperature now gets 
a dependence on the chemical potential. It turns out that an appropriate BW 
potential can be constructed for which both temperature and chemical potential acts
as independent tuning parameters. Exact instanton and the bounce solutions  
are easily computable. While computing these, we find that decay rate from AdS to the black holes phase, 
perhaps not surprisingly, increases with the chemical potential. Introduction of charge, in the
boundary, is represented by turning on the R-charge of ${\cal N} = 4$ SYM. As we
discuss later, in the grand canonical ensemble, various qualitative features of this
gauge theory with non-zero R-charge can still be incorporated into the earlier matrix model 
provided we make appropriate modifications. We find that the instanton and bounce solutions across the
de-confining transition within this modified model can be semi-analytically calculated. Subsequently,
our analysis also shows that, as expected from the bulk considerations, there is an enhancement
of the decay rate with the chemical potential.

The paper is structured as follows. In the second section, we consider the
gauge theory with zero chemical potential. In the bulk, we introduce the BW potential and 
discuss various kink and bounce solutions across the two phases and compute the decay rate.
After a brief summary of the matrix model of \cite{AlvarezGaume:2005fv}, we then compute the kink, bounce solutions
and decay rate around the de-confining transition. Section three addresses similar
issues in the presence of a chemical potential in the grand canonical ensemble. 
Here we minimally modify the above matrix model to incorporate
the effects due to the chemical potential. Once this is done, we go on to check other aspects of this model. In particular, we find that, as predicted from the bulk analysis, the
decay rate from confining to the de-confining phase increases with the chemical potential.

\section{Black hole and their matrix dual: zero chemical potential}

In what follows, we first consider the 
gravity side, analytically compute instanton and bounce at and above the HP temperature. We use this to
compute the value of the effective action on the solution. This provides us with the decay rate 
in the bulk. We then briefly review its phenomenological matrix dual and carry out similar
computations across the de-confinement transition within this matrix model.

\subsection{Bulk}

\begin{figure}[H]
 \centering

 \subfloat[]{
 \includegraphics[width=0.45\textwidth]{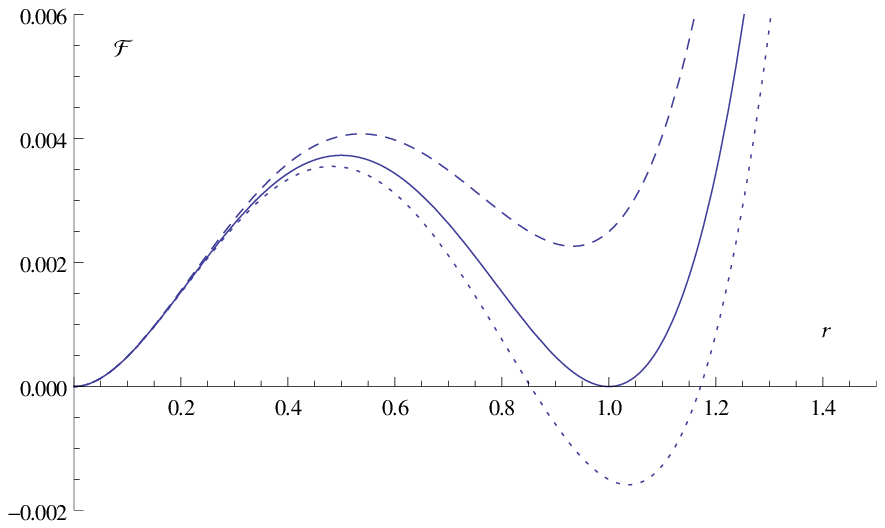}} ~~~~~~~
 \subfloat[]{ 
 \includegraphics[width=0.45\textwidth]{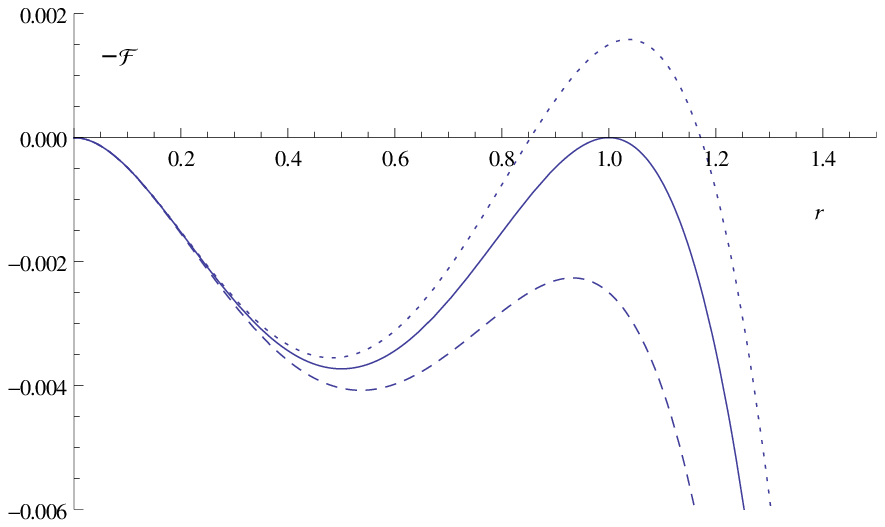}}

 \caption{
{(a) This is ${\cal F}(r, T)$ plotted as a function of $r$. The dashed, solid and dotted
lines are for $T < T_1, T = T_1$ and $T > T_1$ respectively. (b) Inverse of the potential
plotted as a function of $r$.}
}
\label{fig:alpha}
\end{figure}

For the  Schwarzschild-AdS black holes, the BW potential is
given by \cite{Banerjee:2010ve}
\begin{equation}
{\cal F} (r, T) = E - T S = 3(r^2 + 1) r^2  - 4 \pi T r^3.
\label{one}
\end{equation}
Here $r$ measures the horizon size of the black hole, while $T, E, S$ are temperature,
energy and entropy of the hole. Note that all the quantities are made dimensionless
by multiplying them with appropriate powers of AdS radius $l$. $l$ does not appear 
explicitly as it has been set to $1$. Furthermore, we have set
the Newton's constant also to unity. $T$ in (\ref{one}) should be understood 
as a tuning parameter whose dependence on $r$ follows from the on-shell equation
\begin{equation}
\frac{\partial {\cal F}}{\partial r} = 0. 
\label{two}
\end{equation}
The thermal AdS potential can be obtained from (\ref{one}) by setting $r =0$. Further,
the first order transition arises when ${\cal F} = 0$ and equation (\ref{two}) is satisfied.
The solution of these two equations are
\begin{equation}
r = 1, ~~~{\rm  and}~~T_1 =\frac{3}{2\pi}.
\end{equation}
In figure (\ref{fig:alpha} a), we have plotted (\ref{one}) for various temperatures around the
transition point. While stable minimum shown by the dotted line represents the stable 
big black hole for $T >T_1$, the stable minimum at $r =0$, shown in the dashed line, represents
thermal AdS.

\begin{figure}[H]
 \centering

 \subfloat[]{
 \includegraphics[width=0.45\textwidth]{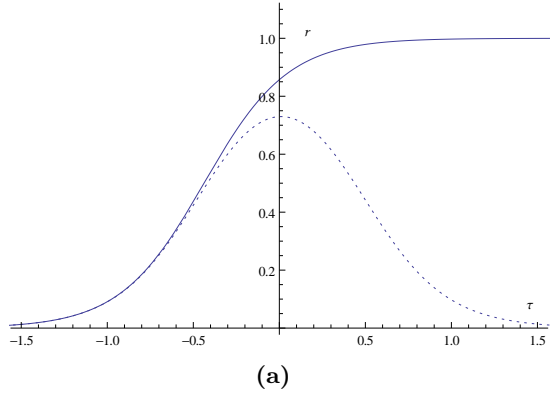}}
 \caption{
{Figure shows the instanton and the bounce solutions. The solid line is for $T = T_1 = 3/(2\pi)$ and the dotted line 
is for $T  = 3.15/(2\pi)$ which is greater than the critical temperature.}
}
\label{fig:betaone}
\end{figure}

Following  \cite{Coleman:1977py}, we need to invert the potential,
and, look for a kink solution connecting degenerate minima at $r = 0$ and $r =1$ at $T = T_1$ shown by the
solid line in (\ref{fig:alpha} b). On the other hand, for $T > T_1$, we need to find the
bounce solution connecting meta-stable and stable minima of the dotted line in (\ref{fig:alpha} b).
This is done by solving the equation
\begin{equation}
\frac{\partial r}{\partial \tau} = {\sqrt{2 {\cal F} (r, T)}},
\end{equation}
where $\tau$ is the Euclidean time. This can be easily integrated with right boundary condition
leading to the solution
\begin{equation}
r = \frac{6 e^{\sqrt{3} \tau}}{1 + 4 \pi T e^{\sqrt{3} \tau} + (4 \pi^2 T^2 -9)e^{2 \sqrt{3} \tau}}. 
\label{three}
\end{equation}
This solution is plotted in figure (\ref{fig:betaone}) for $T \ge T_1$.
For $T= T_1$, it gives a kink solution connecting the minima at $r = 0$ and $r = 1$
representing the thermal AdS and the black hole phase. At a temperature above $T_1$, we 
get a bounce corresponding to the semi-classical tunneling from the metastable AdS phase 
to the stable black hole phase. 

It is straightforward to compute the Euclidean action on (\ref{three}). We get
\begin{eqnarray}
S_E &&= \int_{0}^{1} {\sqrt{ {2 \cal F}(r, T)}} dr \nonumber\\
&&= {\sqrt 2} \Big(\frac{-9 + 6 \pi^2 T^2 + \pi T(-9 + 4 \pi^2 T^2) ~{\rm log}~(3 - 2 \pi T)}
{9 \sqrt 3} \nonumber\\
&&- \frac{1}{27} (-3 + 2 \pi T) [3 {\sqrt {6 - 4\pi T}}(2 + \pi T) \nonumber\\
&&+ {\sqrt 3} \pi T (3 + 2 \pi T) ~{\rm log}~(3 - 2 \pi T + {\sqrt 6} {\sqrt {3 - 2 \pi T}})]\Big).
\end{eqnarray}
For $T = T_1$, the Action evaluates to
\begin{equation}
S_E = \frac{1}{{\sqrt{6}}}.
\end{equation}
The tunneling rate is therefore,
\begin{equation}
\Gamma \sim e^{- \frac{1}{\sqrt{6}\hbar}}.
\end{equation}

\subsection{Boundary}

Having analyzed semi-classical decay of thermal AdS to a stable black hole 
 at high temperature, we now turn our attention to the boundary dual. As mentioned
previously, the framework within which we will search for similar kink and bounce
solutions is the one proposed in \cite{AlvarezGaume:2005fv}\footnote{This model was further
studied and extended in \cite{Basu:2005pj, Dey:2006ds, Dey:2007vt, Dey:2008bw}.} 
. It is a phenomenological 
model where 
partition function of YM theory was expressed as  a matrix integral over the effective action
involving Wilson-Polyakov loop $~({\rm tr}U)/N$. It contains two parameters $a$ and $b$ which, at
fixed 't Hooft coupling, depend on the temperature. The partition function is given by
\begin{equation}
Z(T) = \int dU \Big[a(T) ~{\rm tr}U ~{\rm tr}U^\dagger + \frac{b(T)}{N^2} (~{\rm tr}U ~{\rm tr}U^\dagger)^2\Big].
\label{mod}
\end{equation}
This model was studied in some detail in the original paper \cite{AlvarezGaume:2005fv} and was further reviewed
in section 5 of \cite{Dey:2006ds}. We will be very brief here in describing the model. The effective potential 
arising from this model can be written in terms of $\rho = ~{\rm tr}U ~{\rm tr}U^\dagger/N^2$ as
\begin{eqnarray}
V(\rho) &&= \frac{1-a}{2} \rho^2 - \frac{b}{2} \rho^4 \qquad {\rm for}~~0\le \rho \le \frac{1}{2}\nonumber\\
 &&= -\frac{a}{2} \rho^2 - \frac{b}{2} \rho^4 - \frac{1}{4}~{\rm log}~[2 (1 - \rho)] +\frac{1}{8} \quad
{\rm for}~~\frac{1}{2} \le \rho \le 1.
\label{matricseffective}
\end{eqnarray}
The saddle point equation is given by
\begin{eqnarray}
a \rho + b \rho^3 &&= \rho ~~{\rm for}~0 \ge \rho \ge \frac{1}{2}\nonumber\\
&&=  \frac{1}{1 - \rho} ~~{\rm for}~\frac{1}{2} \ge \rho \ge 1.
\label{sadle}
\end{eqnarray}
The restrictions on $a$ and $b$ are $a < 1, b >0$. We note that the comparison between this
model and the bulk action is valid only when one neglects the string loop effects. The temperature
at which the supergravity description breaks down is identified with the Gross-Witten
transition in the matrix model.

\begin{figure}[t]
\begin{center}
\begin{psfrags}
\psfrag{at}[][]{$a(T)$}
\psfrag{bt}[][]{$b(T)$}
\psfrag{t}[][]{$T$}
\epsfig{file=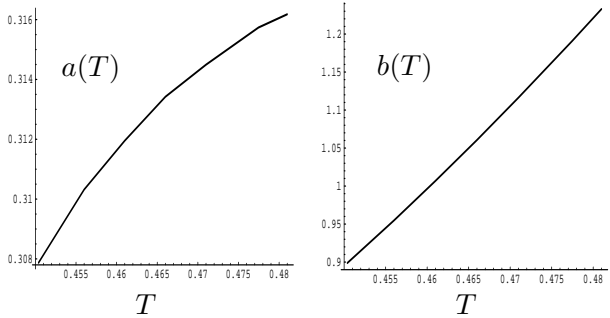, width= 8cm,angle=0}
\end{psfrags}
\vspace{ .1 in }
\caption{Plots of $a(T,0)$ and $b(T,0)$. This is taken from \cite{Dey:2006ds}}
\label{atbt}
\end{center}
\end{figure}

The temperature dependence of the parameters can be fixed as follows. Let us consider 
a temperature $T  > T_0$, $T_0$ being the black hole pair-nucleation temperature.
At this temperature, we can write\footnote{For all the numerical calculations, we set $l$, the
Newton's constant, volume of the three sphere to $1$.}
\begin{equation}
2 a \rho_{1,2}^2 + 2 b \rho_{1,2}^4 + ~{\rm log}~(1 - \rho_{1,2}) + f = - I_{1,2},
\label{i12}
\end{equation}
where $I_{1,2}$ are the actions for the large and the small black holes. $\rho_{1,2}$ 
are the corresponding solutions in the matrix model. The constant $f = ~{\rm log}~(2) - 1/2 $
is introduced to make the potential continuous at $\rho = 1/2$. We further note that
$\rho_{1,2}$ also have to satisfy the saddle point equations (\ref{sadle}). These four
equations uniquely determine $a, b$ and $\rho_{1,2}$ at a specific temperature.
Further, continuing this process for various other temperatures, we get the dependence
of temperature on $a(T)$ and $b(T)$. This can be done numerically and the result is shown is shown in figure
(\ref{atbt}).

\begin{figure}[t]
 \centering

\subfloat[]{
 \includegraphics[width=3in,height=2.5in]{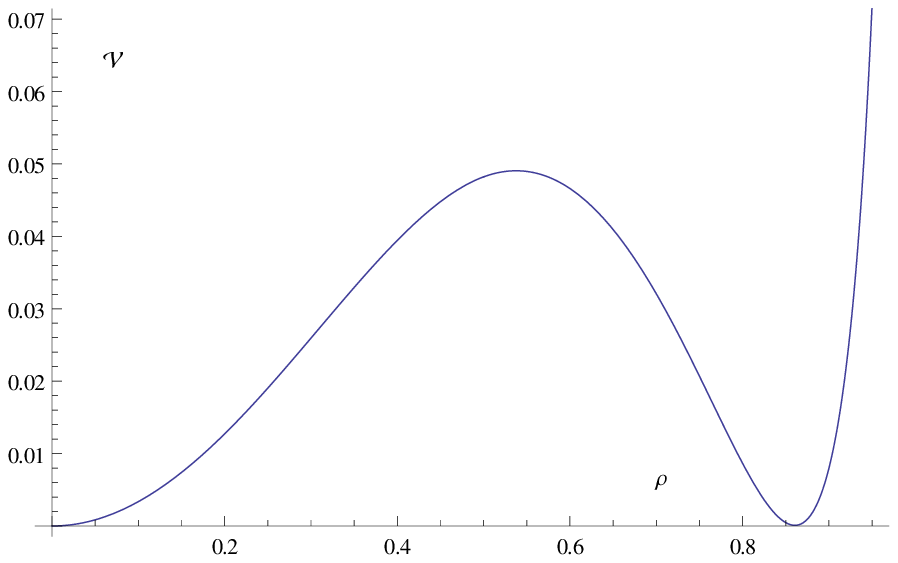}} ~~~~~~~
 \subfloat[]{
 \includegraphics[width=3in,height=2.5in]{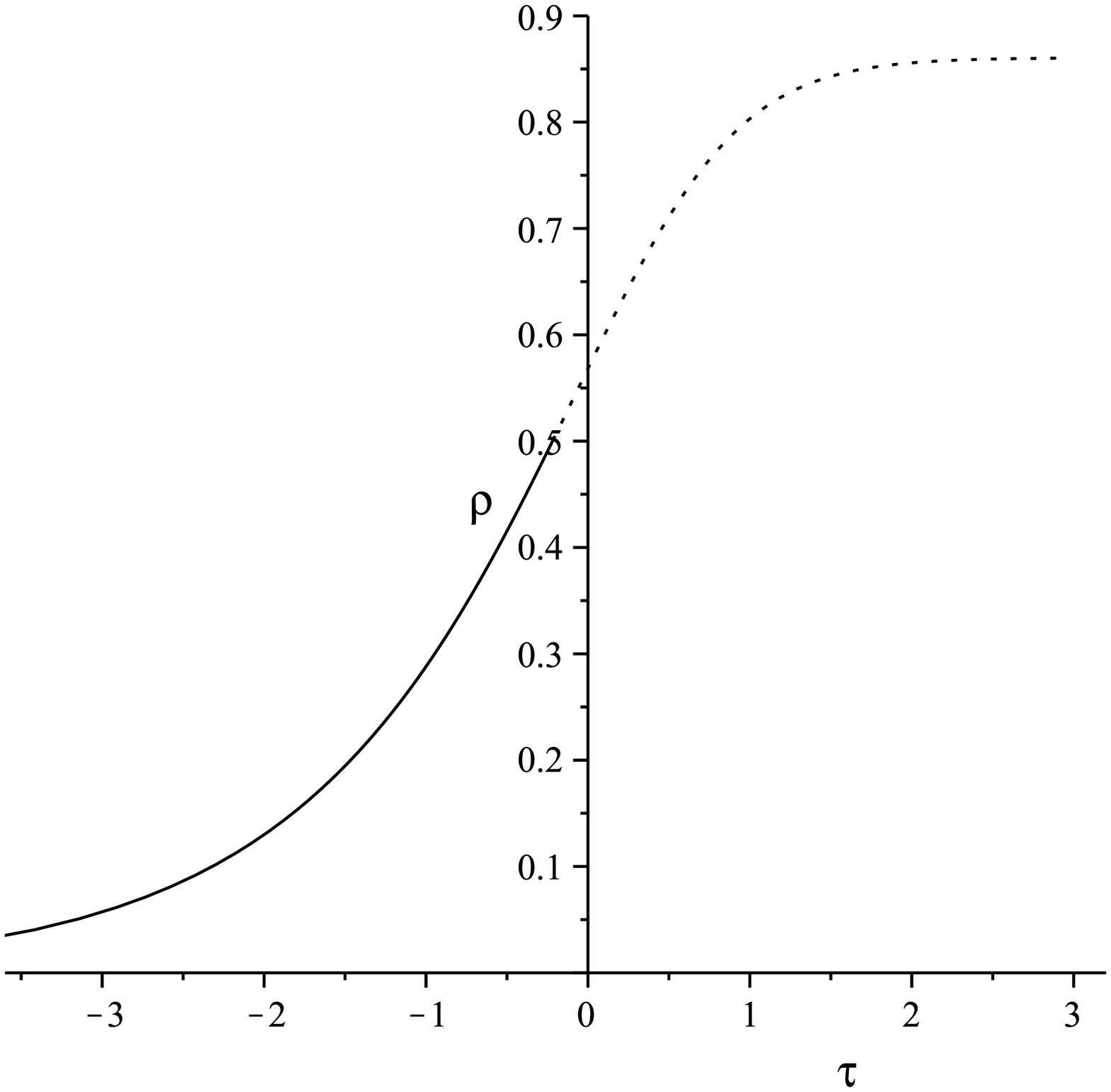}}

 \caption{
{(a) $V(\rho)$ at $T = T_1$ for which $a = 0.316, b = 1.193$. Degenerate minima appear at $\rho =0$ and 
$\rho = 0.856$. (b) Instanton connecting the minima.}}
\label{fig:gamma}
\end{figure}

We now proceed to look at instanton and bounce solution of this model.
Though the method is same as the previous one, in order to proceed, we face a problem. The minimum at non-zero 
$\rho$ always appears 
in a region $\rho > 1/2$. At the de-confining temperature $T = T_1 = 3/(2\pi)$ this minimum appears at
$\rho = \rho_1 = 0.856, a = 0.316, b = 1.193$ and degenerates 
with the one  at $\rho = 0$\footnote{These values follow from  \cite{Dey:2006ds}; see figure 8 there.}
. So to get the complete instanton solution, we need to include integration of
$V$ in (\ref{matricseffective}) in the second region. However, because of the presence of the log
term, we are unable to do an exact integration. In this region, we therefore resort to numerical method.
In the region $0 \le \rho \le 1/2$, the kink solution is given by
\begin{equation}
\rho(\tau) = \frac{ 4 (1 -a) e^{\sqrt{1-a} ~\tau}}{4  + b\,(1-a)\, e^{2 \sqrt{1-a} ~\tau}} ~~~{\rm for}~ 0\le \rho \le 
\frac{1}{2},
\end{equation}
for $a = 0.316, b = 1.193$. For the rest of the region, we numerically integrate the potential.
The result is shown in figure (\ref{fig:gamma}).
For $T=T_1$, $a=0.316$ and $b=1.193$ the critical point is at $\rho=\rho_{0}=0.856$.
Similar to the bulk case a calculation of the action for the instanton solution $\rho$ will give us the width of 
tunneling between the two solutions. A numerical integration of the instanton action evaluates to 
\begin{equation}
\int^{\rho_{0}}_{0}\sqrt{2V(\rho)}d\rho=0.175
\end{equation} 
Hence the tunneling rate becomes
\begin{equation}
\Gamma = e^{-\frac{0.175}{\hbar}}.
\end{equation}

\begin{figure}[H]
 \centering

 \subfloat[]{
 \includegraphics[width=0.43\textwidth]{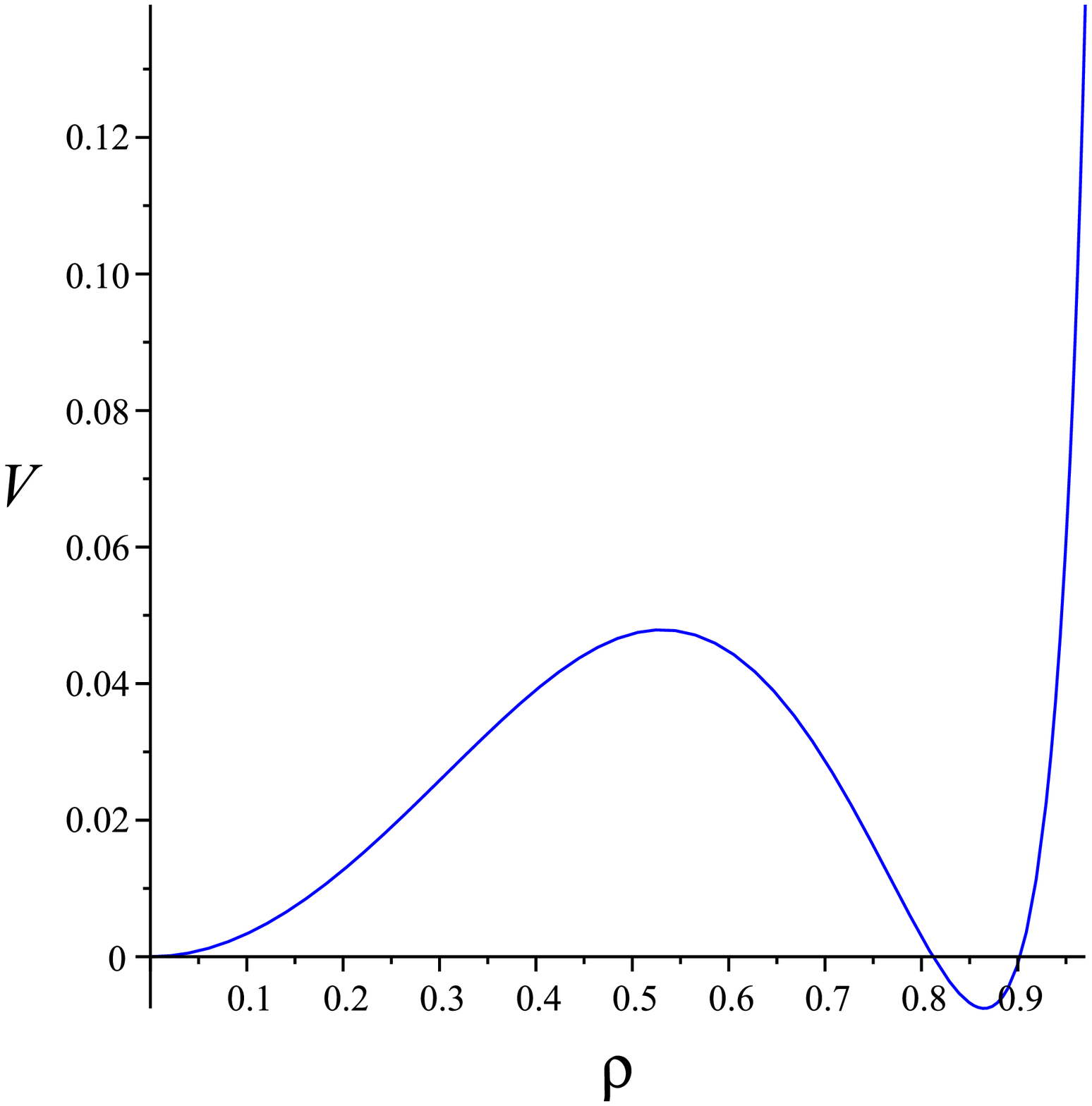}} ~~~~~~~
 \subfloat[]{ 
 \includegraphics[width=0.43\textwidth]{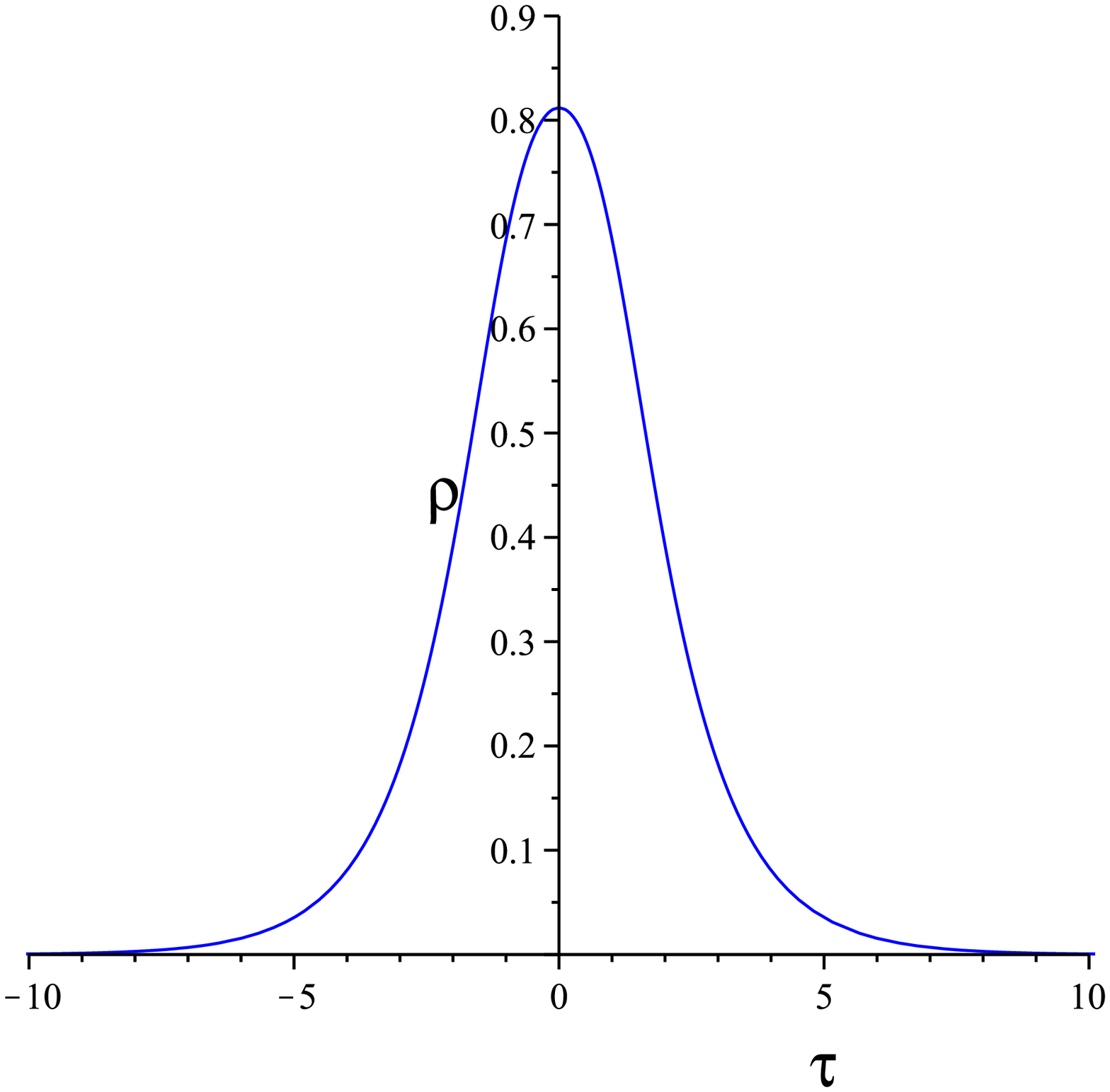}}

 \caption{
{(a) This is $V(\rho)$ plotted as a function of $\rho$ for $T > T_1$,  $a=0.316$ and $b=1.220$.  (b) The bounce solution 
for the same potential with the turning point $\rho_{0}=0.810$}
}
\label{fig:lambda}
\end{figure}

For $T\geq T_1$  the $\rho=0$ solution becomes metastable and, similar to the bulk case, one obtain bounce solutions. 
This is shown in figure (\ref{fig:lambda}).

\section{Black hole and their matrix dual: non-zero chemical potential}

In this section, we will continue similar exploration in the presence of non-zero chemical potential.
Our whole analysis will be carried in the grand canonical ensemble. We first study this problem
in the bulk. In the boundary, we generalize the matrix model to include the effect of chemical potential.
We then find the kink and bounces in this model. Calculating the decay rate from confining
to the de-confining phase, we show that rate increases with the chemical potential. This is
consistent with the expectation that arise from the bulk behavior.

\subsection{Bulk}
In the presence of electrical charge, we need to consider the Reissner-Nordstr\"om black holes.
The BW free energy is easily computed and is given by \cite{Banerjee:2010ve}
\begin{eqnarray}
{\cal F} (r, T, \mu) &=& E - TS - Q \mu\nonumber\\
&=& 3 r^2 (1 - \frac{4}{3} \mu^2 ) - 4 \pi r^3 T + 3 r^4.
\label{rnbw}
\end{eqnarray}
where $Q$ is the charge and $\mu$ is its conjugate chemical potential.
We will restrict ourselves to the case where $\mu \le {\sqrt{3}}/2$.
At the saddle point,
\begin{equation}
r = \frac{3 \pi T + {\sqrt{ 9 \pi^2 T^2 - 3 (6 - 8 \mu^2)}}}{6}.
\end{equation}
This can be inverted to get black hole temperature.
While reducing the temperature below a critical value ($T_1$), this hole becomes globally unstable
and system crosses over to the thermal AdS phase. This critical temperature depends upon the
chemical potential and is given by
\begin{equation}
T_1 = \frac{3 {\sqrt{1 - \frac{4}{3} \mu^2}}}{2\pi}.
\label{mtemp}
\end{equation}

\begin{figure}[H]
 \centering

 \subfloat[]{
 \includegraphics[width=0.45\textwidth]{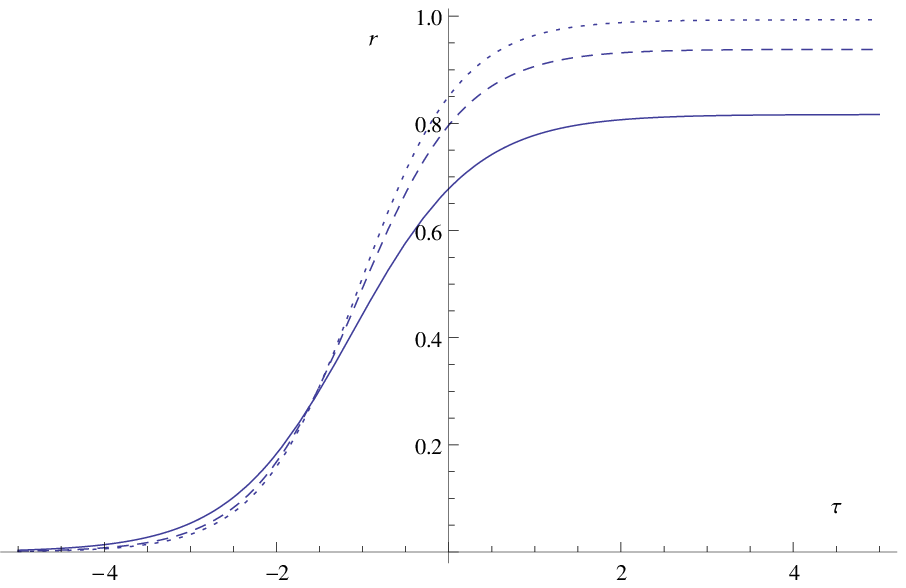}} ~~~~~~~
 \subfloat[]{   
 \includegraphics[width=0.45\textwidth]{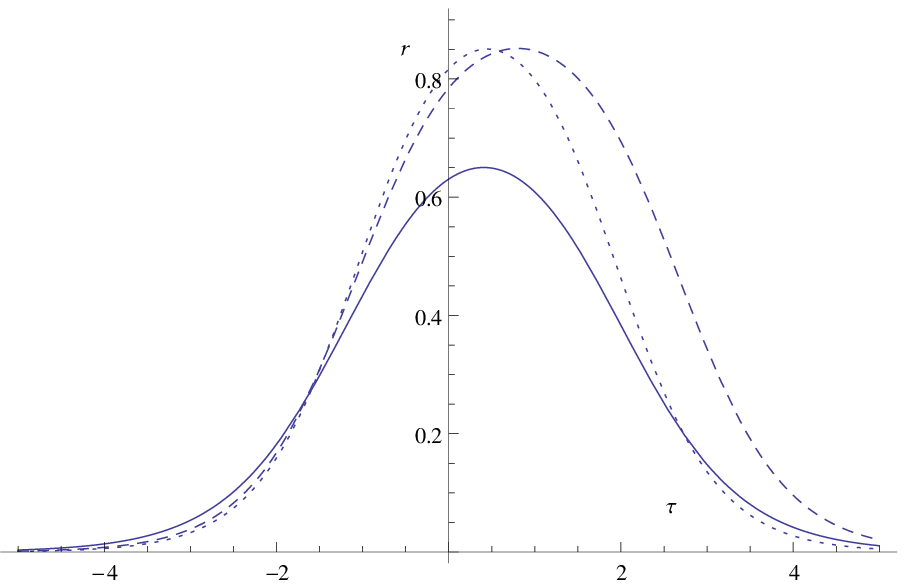}}

 \caption{
{(a) Instanton solution at $T = T_1$ for $\mu = .5, .3$ and $.1$ are shown by solid, dashed and dotted lines 
respectively 
(b) Bounce solution for $T > T_1$ for different $\mu$. Solid, dashed and dotted lines are for 
$\mu = .5, .3$ and $.1$ at temperatures $.4, .45$ and $.48$ respectively. $T_1$ for these $\mu$ values are
$0.390, 0.447, 0.474$ respectively.}
}
\label{fig:delta}
\end{figure}

As in the previous section, the kink solution within this effective description 
can be constructed at the critical temperature. It is given by
\begin{equation} 
r(\tau) = \frac{ 2 (3 - 4 \mu^2) e^{{\sqrt{3 - 4\mu^2}}~\tau}} { 1 + 2 {\sqrt{9 - 12 \mu^2}}
e^{{\sqrt{3 - 4\mu^2}}~\tau}}.
\label{muinst}
\end{equation}
That the solution indeed interpolates between two phases, namely, the thermal AdS and the black hole, 
can be seen from the figure (\ref{fig:delta}a). For $T > T_1$, the thermal AdS phase becomes metastable and we 
expect to find a bounce solution connecting this metastable phase to the stable one representing
the black hole. Indeed this bounce can be explicitly constructed and is given by
\begin{equation}
r (\tau) = \frac{  2 (3 - 4 \mu^2) e^{{\sqrt{3 - 4\mu^2}}~\tau}}
{ 1 + 4 e^{{\sqrt{3 - 4\mu^2}}~\tau} \pi T + e^{2{\sqrt{3 - 4\mu^2}}~\tau} ( 12 \mu^2 + 4 \pi^2 T^2 - 9)}.
\end{equation}
This is plotted for different values of $\mu$ in (\ref{fig:delta}b).
It is straightforward to compute the Euclidean action on the instanton solution (\ref{muinst}). We get
\begin{eqnarray}
&&S_E  = \int_0^{{\sqrt{1 - \frac{4\mu^2}{3}}}  } {\sqrt{2 {\cal F} (r, T, \mu)}}  
\nonumber \\
=&&\sqrt{2}(\frac{1}{27}(3 \sqrt{3-4 \mu^2}
(-3+2 \pi^2 T^2 +4 \mu^2) \nonumber \\
&&+\sqrt{3}\pi  T 
(-9+4 \pi^2 T^2 12\mu^2 ) Log[-2 \pi T+\sqrt{9-12 \mu^2}])  \nonumber \\
&& + (\sqrt{2} (-3+4 \mu^2) (-12 \pi^3 T^3+4 \pi^2 T^2
\sqrt{9-12 \mu^2}+15 \pi  T (3-4 \mu^2) \nonumber \\
&&+6 \sqrt{9-12 \mu^2} (-3+4 
\mu^2)) +\pi  T \sqrt{9-12 \mu^2} (9-4 
 \pi^2 T^2-12 \mu^2) \sqrt{9-12 \mu^2 
-2 \pi  T 
 \sqrt{9-12 \mu^2}} \nonumber \\
&& \log(-2 \pi  T+\sqrt{9-12 \mu^2}+\sqrt{18-24 \mu^2 
- 4\pi  T \sqrt{9-12 \mu^2}})) \nonumber \\
&&/(27 \sqrt{\left(-3+4 \mu^2\right) \left(-9+12 
\mu ^2+2 \pi  T \sqrt{9-12 \mu ^2}\right)}))
\end{eqnarray}
At $T = T_1$ given in (\ref{mtemp}), we get
\begin{equation}
S_E = \frac{(3 - 4 \mu^2)^{\frac{3}{2}}}{9 \sqrt{2}}
\end{equation}  
The tunneling rate is therefore,
\begin{equation}
\Gamma \sim e^{-\frac{(3 - 4 \mu^2)^{\frac{3}{2}}}{9 \sqrt{2} \hbar}}
\end{equation}
We see that decay rate increases with $\mu$.

\subsection{Boundary}

\begin{figure}[H]
 \centering

 \subfloat[]{
 \includegraphics[width=0.45\textwidth]{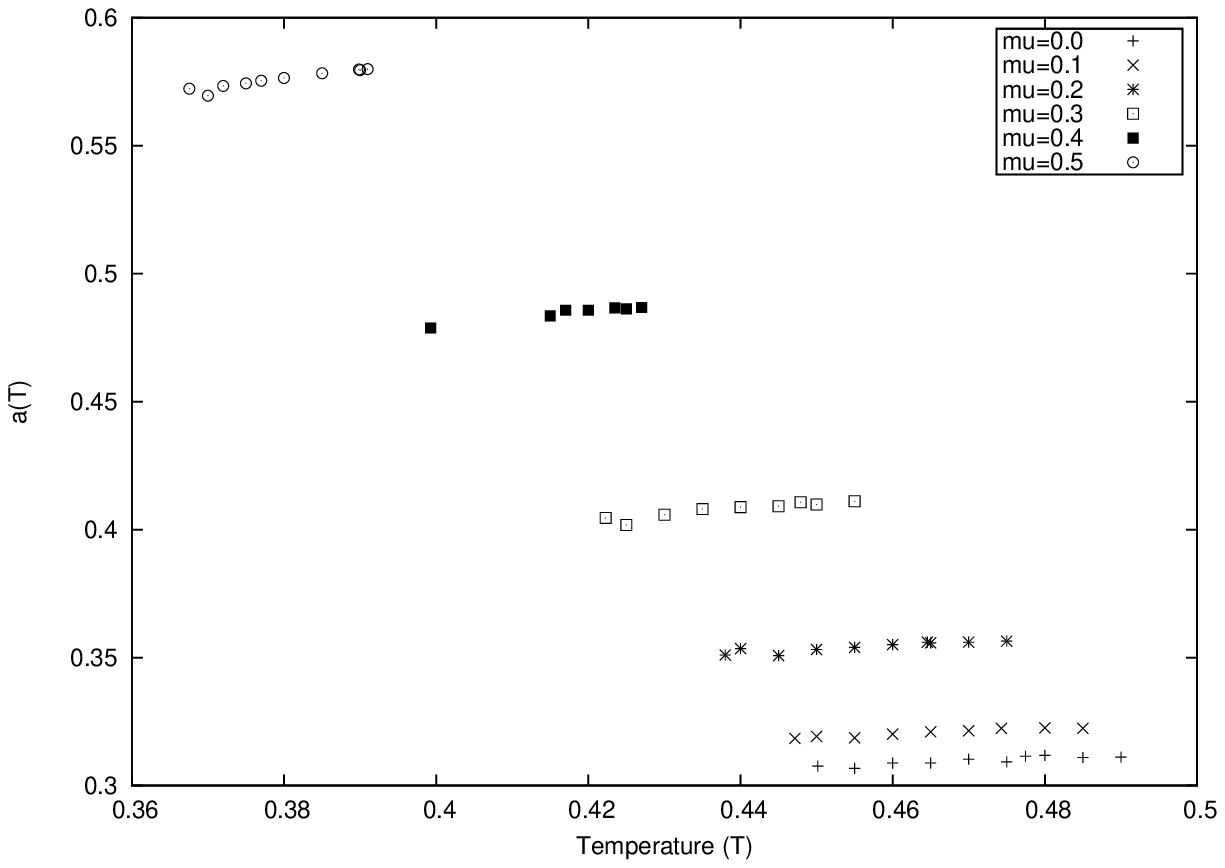}} ~~~~~~~
 \subfloat[]{   
 \includegraphics[width=0.45\textwidth]{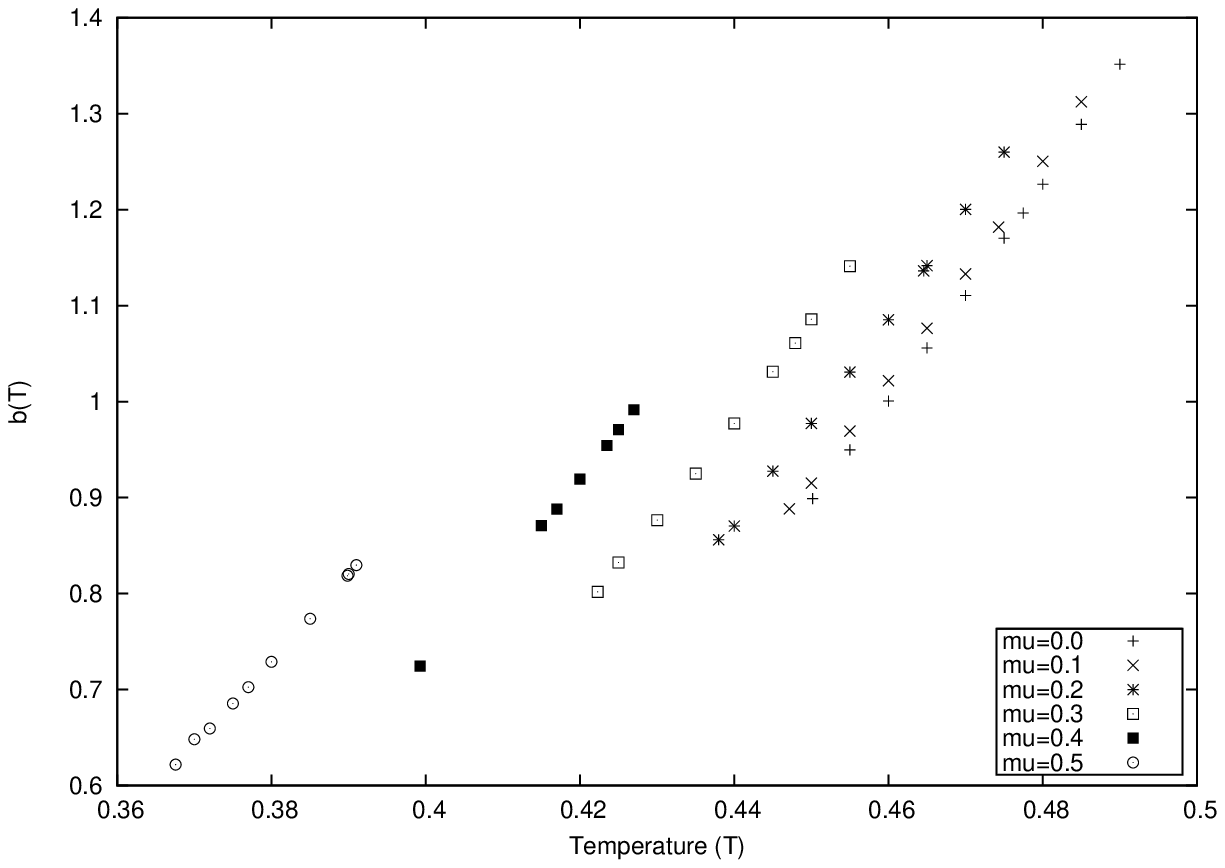}}

 \caption{
{Plots show the behaviours of $a(\mu, T)$ and $b(\mu,T)$.}
}
\label{amubmut}
\end{figure}

We now generalize the  matrix model considered in the previous section to
include the effect of the chemical potential. It turns out that 
minimal modification requires one to keep the structure of the
potential same as (\ref{mod}), but to make $a$ and $b$ dependent
on temperature as well as the chemical potential. One can now follow
the same strategy as before to figure out the dependence of $a, b$
on $\mu$ and $T$. The only difference is that now in  (\ref{i12}),
we need to substitute $I_{1,2}$ for the Reissner-Nordstr\"om black hole.
Carrying out the computation numerically we get $a, b$ as shown in the
figure (\ref{amubmut}). A clear pattern that we see from here is that
both these two parameters increase as we increase the chemical potential.

Once we have the data, we can go ahead and numerically compute the instanton, bounce and the
decay rate. This is similar to the previous section with a little more complicacy
due to non-zero $\mu$. We skip the details and show the results of our computation in the
following figures.

\begin{figure}[H]
 \centering

 \subfloat[]{
 \includegraphics[width=0.45\textwidth]{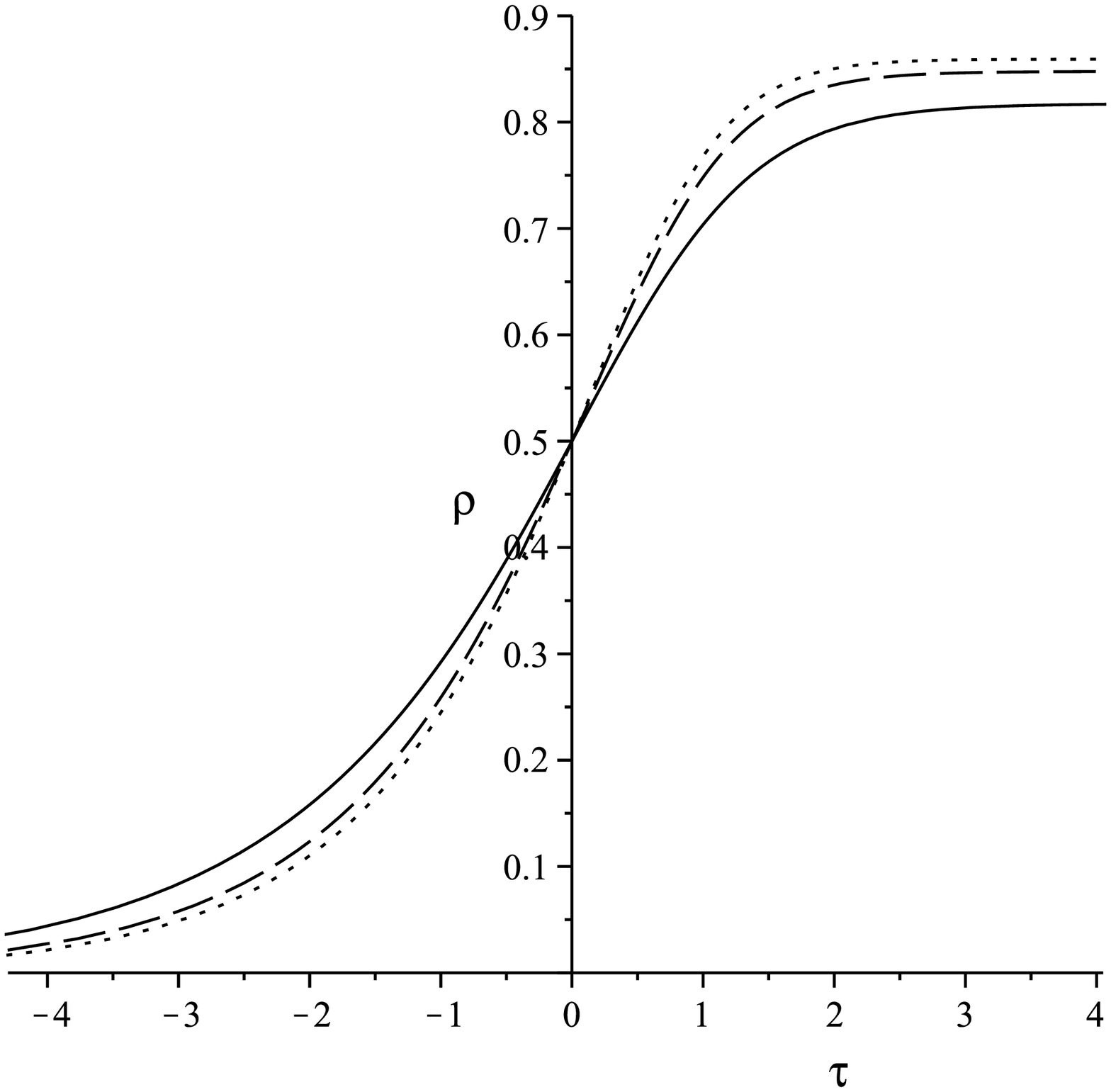}} ~~~~~~~
 \subfloat[]{   
 \includegraphics[width=0.45\textwidth]{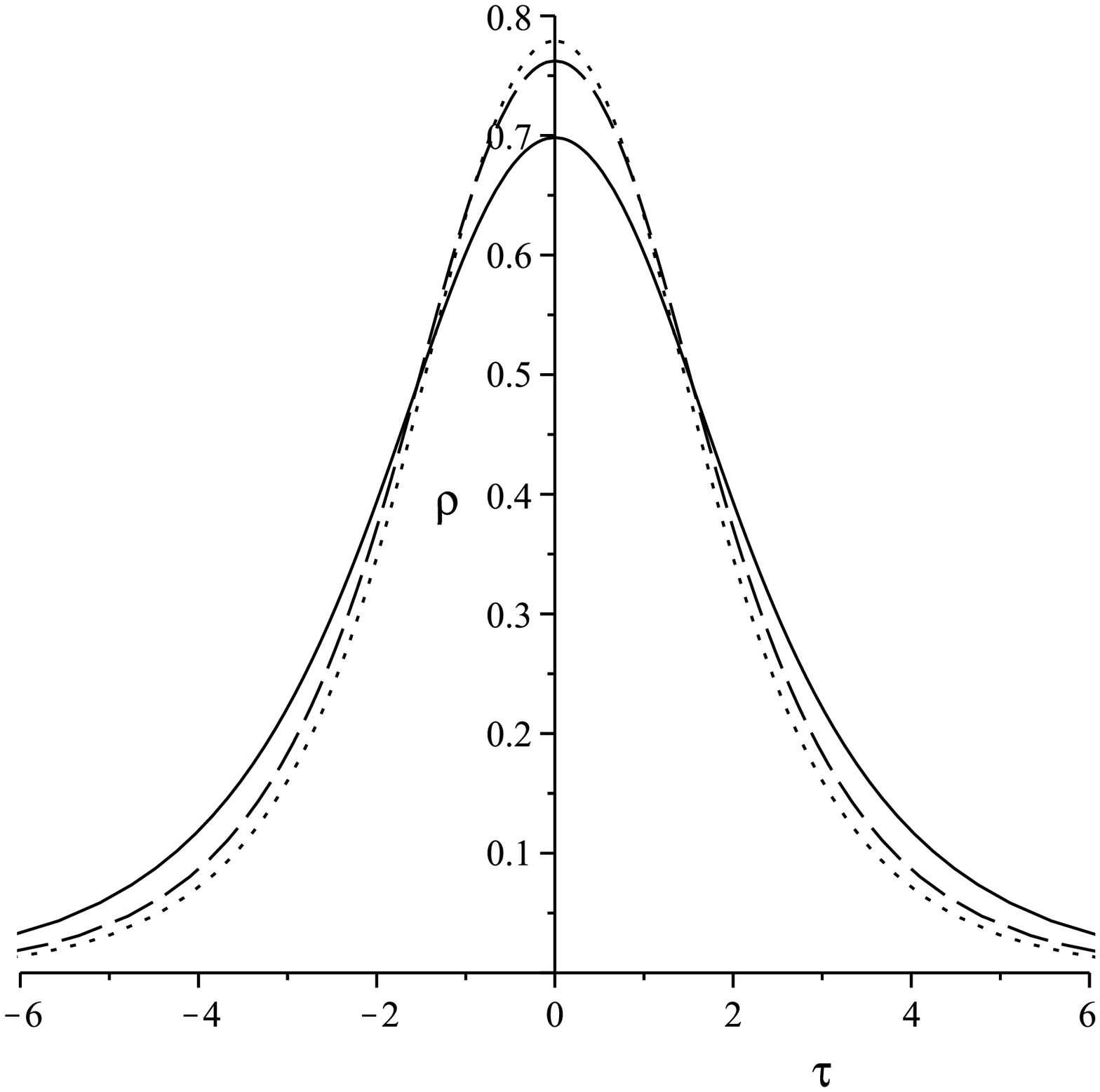}}

 \caption{
{Boundary instanton and bounce curves. Compare with the previous figure
(\ref{fig:delta}) for the bulk case.(a) Solid, dashed and dotted curves
correspond to instanton solutions for $(\mu,T_1)=
(0.5,0.390)\,(0.3,0.447)\,(0.1,0.474)$ respectively. (b) Bounce solutions
at $T>T_1$ for different potentials. The solid, dashed and dotted curves
correspond to $(\mu,T)=(0.5,0.4),\,(0.3,0.454),\,(0.1,0.48)$ respectively.}
}
\label{bdryrN}
\end{figure}

The effective action on the instanton solution can be evaluated numerically for this matrix
model at different chemical potentials. The result is shown in figure (9). It is satisfying to
find that, as in the bluk computation, the value of boundary effective action decreases with the
increase of the chemical potential. This leads to an increase in the tunneling rate. We also note that
the boundray dual is a phenomenonological model and is expected to capture only the qualitative  
features of the gravity dual. The functional forms of ${\cal F}(r)$ and $V(\rho)$ being different, 
it is not surprising that the two lines in figure (9) do not coincide. 

In our study, we have adopted
an effective approach towards the bulk gravity. Though this approach is particularly simple, enabling
us to analytically construct kinks and bounce solutions, it would be instructive to directly
analyze the gravitational bounce, compute the action against the bounce and then compare the results
with the matrix model results. We leave this for our future study. 

\begin{figure}[t]  
\begin{center}
\begin{psfrags}
\epsfig{file=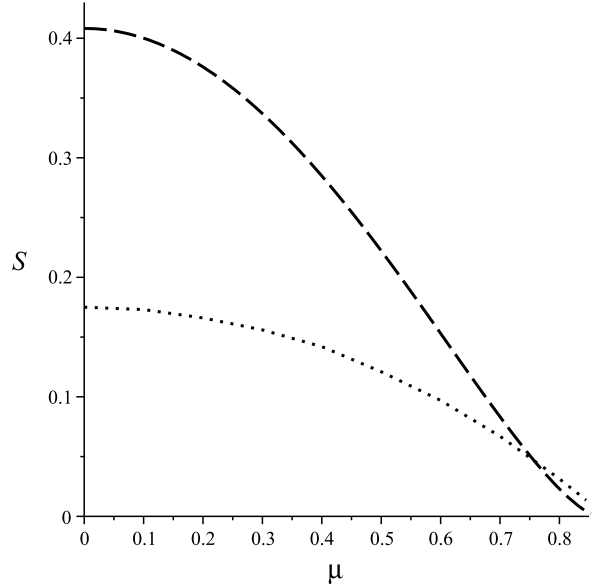, width= 8cm,angle=0}
\end{psfrags}
\vspace{ .1 in }
\caption{Behavior of effective matrix model potential calculated on the instanton is shown by
the dotted line for different chemical potential. For comparison, the corresponding bulk action is
also shown in the plot (dashed line). In both cases, actions decrease with $\mu$.}
\label{fig:betatwo}
\end{center}
\end{figure}

\newpage

\noindent{\bf{Acknowledgments:}} B.C. would like to thank Institute of
Physics, Bhubaneswar for hospitality during the course of this work.

\end{document}